\documentclass[dvips,aoas,preprint]{imsart}

\RequirePackage[OT1]{fontenc}
\RequirePackage{amsthm,amsmath,amssymb}
\RequirePackage[numbers]{natbib}
\RequirePackage[colorlinks,citecolor=blue,urlcolor=blue]{hyperref}

\usepackage{graphicx}
\usepackage{enumerate}

\arxiv{math.PR/0000000}

\startlocaldefs
\numberwithin{equation}{section}
\theoremstyle{definition}
\newtheorem{example}{Example}
\newtheorem*{exrev}{Example 2, revisited}

\def\Pr{{\ensuremath{\rm P}}}%
\def\Exp{{\ensuremath{\rm E}}}%
\def\argmax{\mathop{{\rm arg\,max}}\limits}%
\endlocaldefs

\begin{document}
\begin{frontmatter}
\title{A Bayesian Statistical Approach for Inference on Static
Origin-Destination Matrices}%\protect\thanksref{T1}}
\runtitle{Bayesian Stat. Approach for Inference on Static
OD Matrices}
%\thankstext{T1}{Footnote to the title with the `thankstext' command.}

\begin{aug}
\author{\fnms{Luis} \snm{Carvalho}\thanksref{t1}\ead[label=e1]{lecarval@math.bu.edu}}
\thankstext{t1}{Supported by NSF grant DMS-1107067.}
\runauthor{L. Carvalho}

\affiliation{Boston University}
\address{Department of Mathematics and Statistics\\
Boston University\\
Boston, Massachusetts 02215\\
\printead{e1}}
\end{aug}

\begin{abstract}
We address the problem of static OD matrix estimation from a formal
statistical viewpoint. We adopt a novel Bayesian framework to develop a class
of models that explicitly cast trip configurations in the study region as
random variables. As a consequence, classical solutions from growth factor,
gravity, and maximum entropy models are identified to specific estimators
under the proposed models. We show that each of these solutions usually
account for only a small fraction of the posterior probability mass in the
ensemble and we then contend that the uncertainty in the inference should be
propagated to later analyses or next-stage models. We also propose
alternative, more robust estimators and devise Markov chain Monte Carlo
sampling schemes to obtain them and perform other types of inference. We
present several examples showcasing the proposed models and approach, and
highlight how other sources of data can be incorporated in the model and
inference in a principled, non-heuristic way.
\end{abstract}

\end{frontmatter}

\section{Introduction}
Consider a study region divided into $n$ zones where trips can occur between
any pair of zones. During a certain time period we observe the number of trips
\emph{originated} at zone $i$, $O_i$, and the number of trips \emph{destined}
to zone $j$, $D_j$, for $i, j = 1, \ldots, n$. Our objective is to estimate
the number of trips $T_{ij}$ from each zone $i$ to each zone $j$---including
intrazonal trips $T_{ii}$---conditional on the $\mathcal{O} = \{O_i\}_{i=1}^n$
and $\mathcal{D} = \{D_j\}_{j=1}^n$. Since the trips $\mathcal{T} =
\{T_{ij}\}_{i,j=1,\ldots,n}$ can be represented by the matrix
\begin{equation}
\label{eq:odmatrix}
M = \left[
\begin{array}{cccc}
T_{11} & T_{12} & \cdots & T_{1n} \\
T_{21} & T_{22} & \cdots & T_{2n} \\
\vdots & \vdots & \ddots & \vdots \\
T_{n1} & T_{n2} & \cdots & T_{nn} \\
\end{array}
\right],
\end{equation}
and we are fixing a time window for the trip realizations, our problem is
usually referred to as \emph{static OD matrix estimation}. We note that the OD
matrix $M$ has restrictions on its row and column margins,
\begin{equation}
\label{eq:constraints}
\begin{split}
\sum_{j = 1}^n T_{ij} &= O_i, \quad i = 1, \ldots, n, \\
\sum_{i = 1}^n T_{ij} &= D_j, \quad j = 1, \ldots, n.
\end{split}
\end{equation}
and thus the estimation is constrained. We also require that
$\sum_{i=1}^n O_i = \sum_{j=1}^n D_j \doteq T$ for consistency.
It should be immediate from this formulation that static OD matrix estimation
is a \emph{contingency table} problem\cite{diaconisgangolli}; our goal here is
to provide a broader treatment from a more applied perspective.

This problem has been studied for many decades in the transportation
literature. The first contributions to its solution adopted a physical
interpretation and assumed $\mathcal{T}$ could be described by a gravitational
law \citep{casey55}: $T_{ij} \propto O_i D_j d_{ij}^{-2}$, where $d_{ij}$ is
the distance between zones $i$ and $j$. This functional relation was later
generalized to include decreasing functions of traveling costs $c_{ij}$
between zones $i$ and $j$, called ``deterrence'' functions:
\begin{equation}
\label{eq:gravity}
T_{ij} \propto O_i D_j d(c_{ij}).
\end{equation}
Common choices for $d$ include exponential linear functions of costs, such as
$d(c_{ij}) = \exp(-\beta c_{ij})$ or
$d(c_{ij}) = \exp(-\beta c_{ij} - \alpha \log c_{ij})$.
These gravity models are \emph{synthetic} models since they do not incorporate
previously observed trip patterns. In contrast, \emph{growth factor} models
regard $\mathcal{T}$ as possible future trip patterns and incorporate previous
observations in a doubly constrained formulation. Let the ``seed'' matrix
$\mathcal{T}_0 = \{t_{ij}\}_{i,j=1,\ldots,n}$ be previous observations from
the same or similar study region. Based on the method proposed by
\citet{furness65}, we assume
\begin{equation}
\label{eq:growth}
T_{ij} = A_i O_i B_j D_j t_{ij},
\end{equation}
where $A_i$ and $B_j$ are ``balancing factors'' that are known up to a
proportionality constant. Furness method defines $\mathcal{T}$ by
iteratively solving for the balancing factors to respect constraints
\eqref{eq:constraints} until convergence.

Both gravity and growth factor models provide estimates for $\mathcal{T}$
based on heuristic, functional arguments. \citet{wilson70,wilson74}
defined a formulation based on entropy maximization that would unify both
previous approaches. If
\[
W(\mathcal{T}) = \frac{T!}{\prod_{i,j} T_{ij}!}
\]
is the number of ``micro'' states associated with ``meso'' state
$\mathcal{T}$, then the trip configuration that maximizes $W$, or equivalently
\[
\log W(\mathcal{T}) - \log T! \approx
-\sum_{i,j} \Bigg( T_{ij} \log T_{ij} - T_{ij} \Bigg),
\]
subject to constraints \eqref{eq:constraints} is a maximum entropy solution.
If instead of $\log W$ we maximize
\[
\log W'(\mathcal{T}, \mathcal{T}_0) =
-\sum_{i,j} \Bigg( T_{ij} \log \frac{T_{ij}}{t_{ij}} - T_{ij} \Bigg)
\]
the solution would coincide with the one provided by the Furness model. By
adding an additional cost constraint, such as
\begin{equation}
\label{eq:costconstT}
\sum_{i,j} c_{ij} T_{ij} = C_T
\end{equation}
we obtain the same estimates from the gravity model with $d(c_{ij}) =
\exp(-\beta c_{ij})$.

We can make two important observations from the maximum entropy approach.
First, we note that the functional expressions for $T_{ij}$ from the gravity
and Furness models can actually be regarded as closed form expressions that
can be used to iteratively obtain solutions to a mathematical program that
maximizes $\log W$ or $\log W'$ subject to certain constraints.
Second, since there are many feasible configurations for $\mathcal{T}$, we can
define weights---in Wilson's case given by $W$---to help us find the best trip
configuration; it is, however, implicit from this formulation that any other
trip pattern but the ``optimal'' is also possible, or even likely, to occur.

In this paper we propose a formulation for the OD matrix estimation problem
where $\mathcal{T}$ is explicitly \emph{random}. As we will show, this
formulation corresponds to a Bayesian statistical approach,
e.g.~\citep{gelmanetal}. Even though our focus will be on exploring the
randomness associated with the trip patterns instead of simply extracting a
single trip pattern through optimization, we show that the maximum entropy
solutions, including the classical gravity and growth model solutions, are
identified with maximum \emph{a posteriori} (MAP) estimates under our setup.
Besides this unifying consequence, Bayesian methods also provide other types
of estimators and, more generally, are able to quantify the uncertainty in
estimation and to propagate it to posterior analyses in a principled,
integrated framework.

% TODO
% - connect to contingency tables
% - modern refs (earlier in the text)
% - highlight that not using link count data

\section{Proposed Model}
\label{sec:model}
Let us say that the trips $\mathcal{T}$ are
$(\mathcal{O},\mathcal{D})$-\emph{consistent}, denoted by
$\mathcal{T} \in C(\mathcal{O}, \mathcal{D})$, if $\mathcal{T}$ satisfies
equations~\eqref{eq:constraints}. That is, we define
\[
C(\mathcal{O}, \mathcal{D}) = \Bigg\{
\tilde{\mathcal{T}} = \{\tilde{T}_{ij}\} \,:\,
\sum_{j = 1}^n \tilde{T}_{ij} = O_i \mbox{~and~}
\sum_{i = 1}^n \tilde{T}_{ij} = D_j
\Bigg\}.
\]

As stated before, we regard $\mathcal{T}$ as \emph{random} while margin trips
$\mathcal{O}$ and $\mathcal{D}$ are observed data. As usual in the fully
Bayesian approach we pursue next, all inferences are driven by
the posterior distribution on $\mathcal{T}$ conditional on data
$\mathcal{O}$ and $\mathcal{D}$,
\[
\Pr(\mathcal{T}\,|\,\mathcal{O},\mathcal{D}) =
\frac{\Pr(\mathcal{O}, \mathcal{D}\,|\,\mathcal{T}) \Pr(\mathcal{T})}
{\sum_{\tilde{\mathcal{T}}} \Pr(\mathcal{O},
\mathcal{D}\,|\,\tilde{\mathcal{T}}) \Pr(\tilde{\mathcal{T}})}.
\]

Let us then consider the simple likelihood
\begin{equation}
\label{eq:lhood}
\Pr(\mathcal{O}, \mathcal{D}\,|\,\mathcal{T}) =
I[\mathcal{T} \in C(\mathcal{O}, \mathcal{D})]
\end{equation}
where $I(\cdot)$ is the indicator function: $I(A) = 1$ if and only if $A$ is
true. By the definition of OD consistency, the likelihood in
equation~\eqref{eq:lhood} just states that the margin trips satisfy
equations~\eqref{eq:constraints}, that is, it is a simple indicator for
$(\mathcal{O}, \mathcal{D})$-consistency.

The randomness in trips $\mathcal{T}$ comes initially from our belief, before
observing any data in the margins, of how the trips are distributed. This
belief is hardly subjective, but often arises from experience on similar
regions and zones; in the next section we discuss how to incorporate knowledge
gathered from small scale studies in the same region. To establish a parallel
to the maximum entropy approach of the previous section, we assume that
$\mathcal{T}$ has a conditional multinomial prior distribution given by
$\mathcal{T}\,|\,T \sim \texttt{MN}(T, \mathbf{p})$, that is,
\[
\Pr(\mathcal{T}\,|\,T) = 
\frac{T!}{\prod_{i,j} T_{ij}!} \prod_{i,j} p_{ij}^{T_{ij}},
\]
where $T$ is the total number of trips in the region and
$\mathbf{p} = \{p_{ij}\}_{i,j=1,\ldots,n}$ with $p_{ij}$ being the proportion
of trips between zones $i$ and $j$. Of course, we require that
$\sum_{i,j} p_{ij} = 1$ and $p_{ij}$ are nonnegative. The hyper-prior
parameter $T$ has an improper non-informative distribution $\Pr(T) \propto 1$,
and so the prior becomes
\begin{equation}
\label{eq:prior}
\begin{split}
\Pr(\mathcal{T}) &= \sum_{T = 0}^{\infty} \Pr(\mathcal{T}\,|\,T) \Pr(T)\\
&= \sum_{T = 0}^{\infty}
\frac{T!}{\prod_{i,j} T_{ij}!} \prod_{i,j} p_{ij}^{T_{ij}}
I\Bigg(\sum_{i,j} T_{ij} = T \Bigg)\\
&= \frac{\Big(\sum_{i,j} T_{ij} \Big)!}
{\prod_{i,j} T_{ij}!} \prod_{i,j} p_{ij}^{T_{ij}}.
\end{split}
\end{equation}

The prior on $\mathcal{T}$ resembles the number of micro states $W$ defined by
Wilson, but with the proportions as extra parameters. The proportions
$\mathbf{p}$ have the important role of convening prior information on the
\emph{structure} of trip distribution in the study area. From a behavioral
perspective, $p_{ij}$ corresponds to the probability of a trip in the system,
out of the total $T$ available, occurring between zones $i$ and $j$; we could,
for example, borrowing from random decision theory, define a multinomial logit
model on each $p_{ij}$ that depends on a set of covariates $\mathbf{x}_{ij}$
for each OD pair such as transport costs, time, and user preferences:
\[
p_{ij} = \frac{\exp(\mathbf{x}_{ij}^T \boldsymbol{\beta})}
{\sum_{k,l=1,\ldots,n} \exp(\mathbf{x}_{kl}^T \boldsymbol{\beta})},
\]
where $\boldsymbol{\beta}$ are known coefficients.

While we are now assuming that $\mathbf{p}$ is known and thus fully specifies
$\Pr(\mathcal{T})$ above, we can further incorporate uncertainty by adding
another level of randomness to the prior parameters to form a hierarchical
model; we postpone such considerations to Section~\ref{sec:hierarchical}.

\subsection{Estimation}
The inference we wish to carry out is driven by our updated belief in
$\mathcal{T}$ after observing $\mathcal{O}$ and $\mathcal{D}$ as summarized by
the posterior distribution
\begin{equation}
\label{eq:posterior}
\begin{split}
\Pr(\mathcal{T}\,|\,\mathcal{O}, \mathcal{D})
&= \frac{\Pr(\mathcal{O}, \mathcal{D}\,|\,\mathcal{T}) \Pr(\mathcal{T})}
{\sum_{\tilde{\mathcal{T}}}
\Pr(\mathcal{O}, \mathcal{D}\,|\,\tilde{\mathcal{T}})
\Pr(\tilde{\mathcal{T}})} \\
&= \frac{I[\mathcal{T} \in C(\mathcal{O}, \mathcal{D})] \Pr(\mathcal{T})}
{\sum_{\tilde{\mathcal{T}} \in C(\mathcal{O}, \mathcal{D})}
\Pr(\tilde{\mathcal{T}})} \\
& \propto
\frac{T!}{\prod_{i,j} T_{ij}!} \prod_{i,j} p_{ij}^{T_{ij}}
I[\mathcal{T} \in C(\mathcal{O}, \mathcal{D})].
\end{split}
\end{equation}

One important consequence of $\mathcal{T} \in C(\mathcal{O}, \mathcal{D})$ in
the posterior above is that the prior parameter $T$ implicitly satisfies
\begin{equation}
\label{eq:selfconsist}
T = \sum_{i,j} T_{ij} = \sum_{i=1}^n O_i = \sum_{j=1}^n D_j,
\end{equation}
that is, $\mathcal{O}$ and $\mathcal{D}$ are self-consistent through
$\mathcal{T}$.

A common estimator for $\mathcal{T}$ is the maximum \emph{a posteriori} (MAP)
estimator, the posterior mode:
\begin{equation*}
\begin{split}
\hat{\mathcal{T}}
&= \argmax_{\mathcal{T}} \Big\{
\log \Pr(\mathcal{T}\,|\,\mathcal{O}, \mathcal{D}) \Big\}\\
&= \argmax_{\mathcal{T} \in C(\mathcal{O}, \mathcal{D})}
\Bigg\{ \sum_{i,j} T_{ij} \log p_{ij} - \log T_{ij}! \Bigg\}\\
& \approx \argmax_{\mathcal{T} \in C(\mathcal{O}, \mathcal{D})}
\Bigg\{ \sum_{i,j} T_{ij} \log p_{ij} -
(T_{ij} \log T_{ij} - T_{ij}) \Bigg\} \\
&= \argmax_{\mathcal{T} \in C(\mathcal{O}, \mathcal{D})}
\Bigg\{ -\sum_{i,j} \Bigg(T_{ij} \log \frac{T_{ij}}{p_{ij}} -
T_{ij} \Bigg) \Bigg\}. \\
\end{split}
\end{equation*}
Note the similarity between the maximand and $\log W'$. It is now
straightforward to show that
\[
\hat{T}_{ij} = A_i O_i B_j D_j p_{ij},
\]
where $A_i$ and $B_j$ are balancing factors. Thus, the MAP estimator is
equivalent to the solution obtained from the Furness method for the maximum
entropy formulation. In fact, if we use a prior seed matrix
$\mathcal{T}_0 = \{t_{ij}\}$ to set $p_{ij} = t_{ij} / \sum_{i,j} t_{ij}$,
the prior proportions, we recover the growth factor solution.

To obtain gravity model solutions we just have to define $\mathbf{p}$ based on
an entropy maximizing principle: we want $\mathbf{p}$ that maximizes the
entropy $\mathcal{H}(\mathbf{p}) = -\sum_{i,j} p_{ij} \log p_{ij}$ possibly
subject to additional constraints on $\mathbf{p}$ other than $\sum_{i,j}
p_{ij} = 1$. Since entropy uniquely measures the amount of uncertainty in a
probability distribution, a maximum entropy assignment is justified as the
only unbiased assumption we can attain under a state of partial knowledge of
the system. As \citet[pg. 10]{wilson70} points out, ``the probability
distribution which maximizes entropy makes the weakest assumption which is
consistent with what is known''. If we then constraint on trip costs by
requiring a fixed mean cost in the region
\begin{equation}
\label{eq:costconst}
\sum_{i,j} c_{ij} p_{ij} = C_p,
\end{equation}
we obtain $p_{ij} \propto \exp(-\beta c_{ij})$, and hence a gravity model with
a familiar exponential deterrence function.

Even though setting $\mathbf{p}$ as above provides the same solution, there is
a subtle but important difference to the original maximum entropy formulation:
in Wilson's model we constraint the trip patterns using~\eqref{eq:costconstT},
effectively reducing the number of feasible trip configurations, while in our
proposed model we only restrict the proportions using~\eqref{eq:costconst} to
redefine the weights on trip patterns. In other words, our feasible space is
still only constrained by~\eqref{eq:constraints}, but we set the proportions
as a structural guide for estimation since the shape of the posterior
distribution on $\mathcal{T}$ depends on $\mathbf{p}$. In this sense, we can
think of~\eqref{eq:costconst} as a ``soft'' constraint. We can argue that such
a formulation is more natural since we can certainly have prior knowledge of
overall transport expenditures in the system while it seems artificial to
establish a rigid cost constraint on the whole study region.

Another good estimator is the posterior mean, defined as
\[
\overline{\mathcal{T}}
= \Exp[\mathcal{T}\,|\,\mathcal{O},\mathcal{D}]
= \sum_{\tilde{\mathcal{T}}} \tilde{\mathcal{T}} \cdot
\Pr(\tilde{\mathcal{T}}\,|\,\mathcal{O}, \mathcal{D}).
\]
The posterior mean is more ``robust'' than the posterior mode since it
averages the uncertainty on trip patterns across all possible
$\mathcal{T}$---weighted by their respective posterior probability mass---as
opposed to simply picking the trip pattern with highest posterior probability.
Moreover, since the posterior mean is a linear combination of feasible trip
patterns, it also satisfies the linear constraints in \eqref{eq:constraints}.
There is, however, one major difficulty in this venue: we need to know
$\Pr(\mathcal{T}\,|\,\mathcal{O}, \mathcal{D})$ for each $\mathcal{T}$.

The main hurdle in evaluating the posterior on $\mathcal{T}$
in~\eqref{eq:posterior} is the normalizing factor
$Z(\mathcal{O}, \mathcal{D}) \doteq
\sum_{\tilde{\mathcal{T}} \in C(\mathcal{O}, \mathcal{D})}
\Pr(\tilde{\mathcal{T}})$. Computing $Z(\mathcal{O}, \mathcal{D})$ requires
summing over all possible pairwise trip assignments that are
$(\mathcal{O},\mathcal{D})$-consistent, a daunting task. Before addressing
this central issue, we offer some motivation in the next subsection.

\subsection{A simple example}
Suppose that, for $n=2$ zones, we observe $O_1$, $O_2$, $D_1$, $D_2$, and
wish to estimate the entries $\mathcal{T}$ in the OD matrix
\[
\begin{array}{cc|c}
T_{11} & T_{12} & O_1 \\
T_{21} & T_{22} & O_2 \\ \hline
D_1 & D_2 & T
\end{array}
\]
with margins and total number of trips $T$ displayed.

Since $\mathcal{T}$ is consistent, we know that $T_{12} = O_1 - T_{11}$,
$T_{21} = D_1 - T_{11}$ and $T_{22} = O_2 - T_{11} = T_{11} - (T - O_2 - D_2)
= T_{11} - \Delta$, where we set $\Delta \doteq T - O_2 - D_2$. The posterior
on $\mathcal{T}$ is then a posterior on $T_{11}$ due to these linear
constraints:
\begin{equation}
\label{eq:postex}
\begin{split}
\Pr(T_{11}\,|\,\mathcal{O}, \mathcal{D}) &\propto
\frac{T!}{T_{11}! T_{12}! T_{21}! T_{22}!}
p_{11}^{T_{11}} p_{12}^{T_{12}} p_{21}^{T_{21}} p_{22}^{T_{22}} \\
&\propto
\frac{p_{11}^{T_{11}} p_{12}^{O_1 - T_{11}} p_{21}^{D_1 - T_{11}}
p_{22}^{T_{11} - \Delta}}
{T_{11}! (O_1 - T_{11})! (D_1 - T_{11})! (T_{11} - \Delta)!} \\
&\propto
\binom{O_1}{T_{11}} \binom{D_1 - \Delta}{D_1 - T_{11}} \psi^{T_{11}} \\
&\doteq H(T_{11}; O_1, D_1, \Delta, \psi), \\
\end{split}
\end{equation}
where $\psi = (p_{11} p_{22}) / (p_{12} p_{21})$ can be interpreted as a
intra-interzonal odds ratio. Since $D_1 - \Delta = T - O_1$,
we can see that $T_{11}$ follows a \emph{non-central hypergeometric
distribution}\cite{mccullaghnelder}:
\[
T_{11}\,|\,\mathcal{O}, \mathcal{D} \sim \texttt{HG}(O_1, D_1, \Delta; \psi).
\]

Note that $\mathcal{T} \in C(\mathcal{O}, \mathcal{D})$ is equivalent to
requiring that $\max\{0, \Delta\} \le T_{11} \le \min\{O_1, D_1\}$, and so the
normalizing constant for \eqref{eq:postex} is the sum of its right-hand side
over the values of $T_{11}$ above. In practice, however, it is simpler to
obtain posterior samples of $T_{11}$ using a \emph{Metropolis-Hastings}
algorithm \citep{hastings70,gilksetal,givenshoeting}.

As proposal we adopt a random walk: given our actual position $T_{11}^{(t-1)}$
at iteration $t-1$, we set our candidate $T_{11}^*$ a step to the left,
$T_{11}^* = T_{11}^{(t-1)} - 1$ with probability $0.5$ or a step to the right,
$T_{11}^* = T_{11}^{(t-1)} + 1$ with probability $0.5$.
If $T_{11}^* < \max\{0,\Delta\}$ or $T_{11}^* > \min\{O_1,D_1\}$ we
immediately reject $T_{11}^*$---and set $T_{11}^{(t)} = T_{11}^{(t-1)}$---as
it is out of bounds. Otherwise we accept $T_{11}^*$---and thus set
$T_{11}^{(t)} = T_{11}^*$---with probability $\min\{R(T_{11}^{(t-1)},
T_{11}^*), 1\}$, where $R(T_{11}^{(t-1)}, T_{11}^*)$ is the acceptance
ratio
\[
R(T_{11}^{(t-1)}, T_{11}^*) =
\frac{H(T_{11}^*; O_1, D_1, \Delta, \psi)}
{H(T_{11}^{(t-1)}; O_1, D_1, \Delta, \psi)}.
\]

We denote this Metropolis step by
\[
T_{11}^{(t)} = MS(T_{11}^{(t-1)}; O_1, D_1, \Delta, \psi).
\]
To summarize, we can obtain samples from $T_{11}$ by doing:
\begin{enumerate}[Step 1.]
\item Start at some arbitrary initial $T_{11}^{(0)}$.
\item For $t = 1, 2, \ldots$ do (until convergence): execute a Metropolis
step,
\[
T_{11}^{(t)} = MS(T_{11}^{(t-1)}; O_1, D_1, \Delta, \psi),
\]
that is,
  \begin{enumerate}[Step 2.1.]
  \item Sample candidate $T_{11}^*$: sample $U \sim U(0,1)$; if $U < 0.5$ set
  $T_{11}^* = T_{11}^{(t-1)} - 1$, otherwise set
  $T_{11}^* = T_{11}^{(t-1)} + 1$.
  \item If $T_{11}^* < \max\{0,\Delta\}$ or $T_{11}^* > \min\{O_1,D_1\}$ set
  $T_{11}^{(t)} = T_{11}^{(t-1)}$ (reject). Otherwise, sample $U \sim U(0,1)$:
  if $U < \min\{R(T_{11}^{(t-1)}, T_{11}^*), 1\}$ then set $T_{11}^{(t)} =
  T_{11}^*$ (accept), else set $T_{11}^{(t)} = T_{11}^{(t-1)}$ (reject).
  \end{enumerate}
\end{enumerate}

A numerical example should help us further gain intuition on the problem.

\begin{example}
Let $O_1 = 40$, $O_2 = 40$, $D_1 = 60$, $D_2 = 20$, $p_{11} = 0.1$, $p_{12} =
0.2$, $p_{21} = 0.3$, and $p_{22} = 0.4$. It follows that $T = O_1 + O_2 = D_1
+ D_2 = 80$, $\Delta = T - O_2 - D_2 = 20$, and
$\psi = (p_{11} p_{22}) / (p_{12} p_{21}) = (0.1 \cdot 0.4) / (0.2 \cdot 0.3)
= 2/3$, and so $T_{11} \sim \texttt{HG}(40, 60, 20; 2/3)$.

Using random walk Metropolis samples $T_{11}^{(1)}, \ldots, T_{11}^{(G)}$ we
can produce point estimates for $T_{11}$ if desired: the posterior mean,
\[
\overline{T}_{11} = \Exp[T_{11}\,|\,\mathcal{O}, \mathcal{D}] \approx
\frac{1}{G} \sum_{g=1}^G T_{11}^{(g)},
\]
and the posterior mode,
\[
\hat{T}_{11} =
\argmax_{x=\max\{0,\Delta\},\ldots,\min\{O_1,D_1\}}
\Pr(T_{11} = x\,|\,\mathcal{O}, \mathcal{D}).
\]
$\hat{T}_{11}$ can be obtained from estimates for
$\Pr(T_{11}\,|\,\mathcal{O}, \mathcal{D})$, by Monte Carlo simulation,
\begin{equation}
\label{eq:mcprob}
\Pr(T_{11} = x\,|\,\mathcal{O}, \mathcal{D}) \approx
\frac{1}{G} \sum_{g=1}^G I(T_{11}^{(g)} = x),
\end{equation}
or from the Furness method. Using $G=10,\!000$, we obtain $\overline{T}_{11} =
28.43$ and $\hat{T}_{11} = 28.49$, and so both the posterior mean and
posterior mode, estimated from our samples and rounded to the nearest feasible
integer, are $\approx 28$. It is not uncommon for both estimates to coincide,
especially when the distribution is unimodal and close to symmetric, as in
this case.

Interestingly,
$\Pr(T_{11} = 28\,|\,\mathcal{O}, \mathcal{D}) \approx 0.20$; even for this
simple example with a small number of trips we can see that the probability of
the most probable trip configuration corresponds to a small fraction of
possible configurations. This effect should not come as a surprise: as the
number of zones and margins grow, so do the number of possible consistent
configurations, and so the probability of any single trip configuration
becomes even smaller.

We have previously remarked on the structural role of the proportions
$\mathbf{p}$, serving as a guide when searching for a representative trip
pattern among the many possible feasible configurations. We note, however,
that there is no principled reason to expect a close relation between
$\mathbf{p}$ and actual proportions $\mathcal{T}/T$ since the latter is
constrained by origin and destination margins. As an example, consider
Figure~\ref{fig:boxplot0}, where we show the marginal posterior distributions
of $T_{11}$, $T_{12}$, $T_{21}$, and $T_{22}$, along with expected
``structural'' number of trips given by $T\mathbf{p}$. The discrepancies are
clear once we observe that $Tp_{11} + Tp_{12} = 24 < 40 = O_1$ and similarly
for the other margins; equivalently,
$(T_{11} + T_{12})/T = 0.5 > 0.3 = p_{11} + p_{12}$ for any (feasible) trip
pattern $\mathcal{T}$.

\begin{figure}[htbp]
\begin{center}
\includegraphics[width=.7\textwidth]{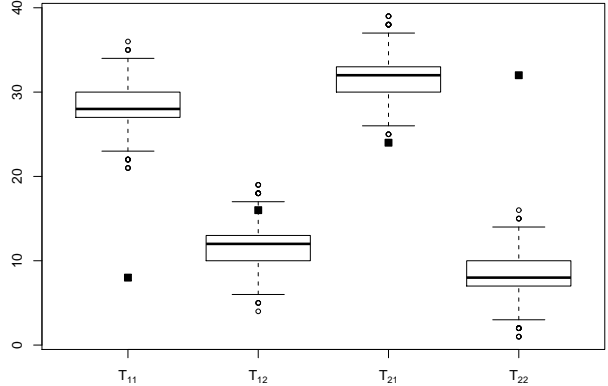}
\end{center}
\caption{Estimated posterior distributions of $\mathcal{T}$ from
$10,\!000$ samples. Squares mark expected structural trips.}
\label{fig:boxplot0}
\end{figure}

\end{example}

\subsection{Posterior sampler}
Let us now extend the results from the last section to our problem. In
general, for $n$ zones we have the following OD matrix with margins displayed:
\[
\begin{array}{cccc|c}
T_{11} & T_{12} & \cdots & T_{1n} & O_1 \\
T_{21} & T_{22} & \cdots & T_{2n} & O_2 \\
\vdots & \vdots & \ddots & \vdots & \vdots \\
T_{n1} & T_{n2} & \cdots & T_{nn} & O_n \\ \hline
D_1 & D_2 & \cdots & D_n & T
\end{array}
\]

We now proceed to eliminate the first $n-1$ entries in the last row and column
by means of the linear constraints in the margins:
\begin{equation}
\label{eq:lastconst}
\begin{split}
T_{nj} = D_j - \sum_{i=1}^{n-1} T_{ij}, \qquad j = 1, \ldots, n-1, \\
T_{in} = O_i - \sum_{j=1}^{n-1} T_{ij}, \qquad i = 1, \ldots, n-1. \\
\end{split}
\end{equation}
The corner entry $T_{nn}$ requires special handling:
\begin{equation}
\label{eq:cornerconst}
\begin{split}
T_{nn} &= O_n - \sum_{j=1}^{n-1} T_{nj} \\
&= O_n - \sum_{j=1}^{n-1} \Bigg( D_j - \sum_{i=1}^{n-1} T_{ij} \Bigg) \\
&= \sum_{i,j=1}^{n-1} T_{ij} - \Bigg( \sum_{j=1}^{n-1} D_j - O_n \Bigg) \\
&= \sum_{i,j=1}^{n-1} T_{ij} - \underbrace{(T - O_n - D_n)}_{\Delta}.
\end{split}
\end{equation}
Ultimately, $T_{nn}$ stems from the symmetry in
equation~\eqref{eq:selfconsist}.

To sample from the entries in the $(n-1)$-by-$(n-1)$ upper submatrix $S$ we
adopt a \emph{Gibbs sampler} \citep{gemangibbs}; see also
\citep{gilksetal,givenshoeting}. The conditional posterior distributions are
$\Pr(T_{ij}\,|\,T_{[ij]}, \mathcal{O}, \mathcal{D})$, for $i,j=1,\ldots,n-1$,
where $T_{[ij]}$ denotes all the entries in $\mathcal{T}$ but $T_{ij}$, that
is, $T_{[ij]} \doteq \{T_{kl}\}_{k,l=1,\ldots,n-1, k\ne i, l\ne j}$. The only
terms in $\Pr(\mathcal{T}\,|\,\mathcal{O}, \mathcal{D})$ that depend on
$T_{ij}$ are now related to $T_{in}$ and $T_{nj}$ through
equations~\eqref{eq:lastconst} and to $T_{nn}$ through
equation~\eqref{eq:cornerconst}. Namely,
\begin{equation}
\begin{split}
\Pr(T_{ij}\,|\,T_{[ij]}, \mathcal{O}, \mathcal{D}) &\propto
\frac{p_{ij}^{T_{ij}} p_{in}^{T_{in}}
p_{nj}^{T_{nj}} p_{nn}^{T_{nn}}}
{T_{ij}! T_{in}! T_{nj}! T_{nn}!} \\
&\doteq
\frac{p_{ij}^{T_{ij}} p_{in}^{O_{ij} - T_{ij}}
p_{nj}^{D_{ij} - T_{ij}} p_{nn}^{T_{ij} - \Delta_{ij}}}
{T_{ij}! (O_{ij} - T_{ij})!
(D_{ij} - T_{ij})! (T_{ij} - \Delta_{ij})!} \\
&\propto
\binom{O_{ij}}{T_{ij}} \binom{D_{ij} - \Delta_{ij}}{D_{ij} - T_{ij}}
\psi_{ij}^{T_{ij}},\\
\end{split}
\end{equation}
where we define
$O_{ij} \doteq
O_i - \sum_{l=1,\ldots,n-1, l\ne j} T_{il}$,
$D_{ij} \doteq
D_j - \sum_{k=1,\ldots,n-1, k \ne i} T_{kj}$,
$\Delta_{ij} \doteq
\Delta - \sum_{k,l=1,\ldots,n-1, k \ne i, l \ne j} T_{kl}$,
and $\psi_{ij} \doteq (p_{ij} p_{nn})/(p_{in} p_{nj})$---a ``within-between''
odds trip ratio---to simplify the expressions. Thus,
\begin{equation}
\label{eq:postcond}
T_{ij}\,|\,T_{[ij]}, \mathcal{O}, \mathcal{D} \sim
\texttt{HG}(O_{ij}, D_{ij}, \Delta_{ij}; \psi_{ij}).
\end{equation}

It is now straightforward to sample from the posterior for $\mathcal{T}$ using
a hybrid Metropolis-within-Gibbs sampling scheme since we know how to sample
from the non-central hypergeometric:

\begin{enumerate}[Step 1.]
\item Start at some arbitrary initial configuration $\mathcal{T}^{(0)}$.
\item For $t = 1, 2, \ldots$ do (until convergence):
  \begin{enumerate}[Step 2.1.]
  \item For $i,j = 1, \ldots, n-1$ do:
  sample
  $T_{ij}^{(t)} \sim T_{ij}\,|\,T_{[ij]}^{(t-1)}, \mathcal{O}, \mathcal{D}$
  in~\eqref{eq:postcond} using a Metropolis step,
  \[
  T_{ij}^{(t)} =
  MS(T_{ij}^{(t-1)}; O_{ij}^{(t-1)}, D_{ij}^{(t-1)}, \Delta_{ij}^{(t-1)},
  \psi_{ij}),
  \]
  with $O_{ij}^{(t-1)}$, $D_{ij}^{(t-1)}$, $\Delta_{ij}^{(t-1)}$, and
  $\psi_{ij}$ defined as above. Note that all the parameters but $\psi_{ij}$
  depend on $T_{[ij]}^{(t-1)}$ and so carry an iteration index.
  \end{enumerate}
\end{enumerate}

It should be noted that this sampling scheme is similar to the more general
scheme from algebraic statistics and based on Markov basis \cite{diaconissturmfels}.

\begin{example}
We end this section with an example taken from
\citep[pg.~179]{ortuzarwillusen}. The costs $\{c_{ij}\}$ between four zones
are listed in Table~\ref{tab:ex1}, along with observed origin and destination
margins.

\begin{table}[bht]
\caption{Trip costs between four zones with observed origin and destination
margins. Reproduced from \citep[table~5.8]{ortuzarwillusen}.}
\label{tab:ex1}
\begin{tabular*}{\textwidth}
{c@{\extracolsep{\fill}}c@{\extracolsep{\fill}}c@{\extracolsep{\fill}}c@{\extracolsep{\fill}}c@{\extracolsep{\fill}}c} \hline
Zone & $\mathbf{1}$ & $\mathbf{2}$ & $\mathbf{3}$ & $\mathbf{4}$ &
$O_i$ \\ \hline
$\mathbf{1}$ & $3$ & $11$ & $18$ & $22$ & $400$ \\
$\mathbf{2}$ & $12$ & $3$ & $13$ & $19$ & $460$ \\
$\mathbf{3}$ & $15.5$ & $13$ & $5$ & $7$ & $400$ \\
$\mathbf{4}$ & $24$ & $18$ & $8$ & $5$ & $702$ \\ \hline
$D_j$ & $260$ & $400$ & $500$ & $802$ & $\mathbf{1962}$ \\ \hline
\end{tabular*}
\end{table}

Let us now assume that $p_{ij} \propto \exp(-\beta c_{ij})$ with
$\beta = 0.10$. After running our Gibbs sampler until assumed convergence, we
take $G = 10,\!000$ samples to perform posterior inference; the marginal
posterior distributions for $T_{ij}$ in the upper $3$-by-$3$ matrix are
summarized in Figure~\ref{fig:boxplot1}.

\begin{figure}[htbp]
\begin{center}
\includegraphics[width=.7\textwidth]{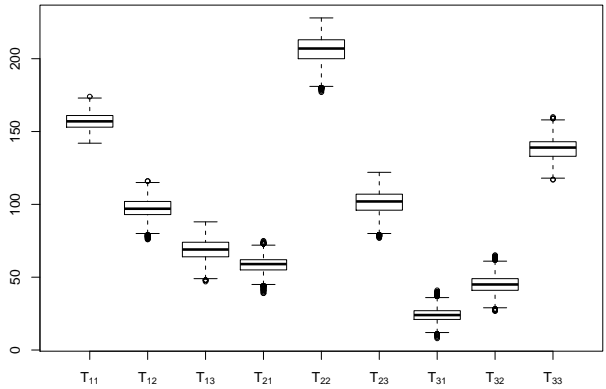}
\end{center}
\caption{Estimated posterior distributions of $\mathcal{T}$ from
$10,\!000$ samples.}
\label{fig:boxplot1}
\end{figure}

The posterior mean $\overline{\mathcal{T}}$, estimated from our samples by
\begin{equation}
\label{eq:posteriormean}
\overline{\mathcal{T}} = \Exp[\mathcal{T}\,|\,\mathcal{O},\mathcal{D}]
\approx \frac{1}{G} \sum_{g=1}^G \mathcal{T}^{(g)}
\end{equation}
is very similar to the Furness solution reported in \citep{ortuzarwillusen}.
We list $\overline{\mathcal{T}}$ along with $95\%$ credible intervals for each
$T_{ij}$ in Table~\ref{tab:post1}. The credible intervals are wider than in
our previous simple example due to the much higher number of feasible
configurations in $C(\mathcal{O}, \mathcal{D})$. In fact, we estimate from the
posterior samples that
$\Pr(\mathcal{T} = \overline{\mathcal{T}}\,|\,\mathcal{O}, \mathcal{D})
\approx \Pr(\mathcal{T} = \hat{\mathcal{T}}\,|\,\mathcal{O}, \mathcal{D})
\approx 2 \cdot 10^{-3}$. Since the most probable trip pattern accounts for
only $0.2\%$ of the posterior probability mass, we can conclude that even the
Furness solution has little support from the data. Interval estimators now
become more attractive representatives of the posterior space of trip
configurations given a desired credibility level.

\begin{table}[bht]
\caption{Posterior mean and $95\%$ credible intervals.}
\label{tab:post1}
\begin{tabular*}{\textwidth}
{c@{\extracolsep{\fill}}c@{\extracolsep{\fill}}c@{\extracolsep{\fill}}c@{\extracolsep{\fill}}c} \hline
Zone & $\mathbf{1}$ & $\mathbf{2}$ & $\mathbf{3}$ & $\mathbf{4}$ \\ \hline
$\mathbf{1}$ & $ 157.14$~~$[147, 169]$ & $  97.37$~~$[ 85, 110]$ &
$  68.73$~~$[ 56,  81]$ & $  76.75$~~$[64, 91]$ \\
$\mathbf{2}$ & $  58.70$~~$[ 48,  68]$ & $ 206.35$~~$[190, 221]$ &
$ 101.27$~~$[ 84, 116]$ & $  93.69$~~$[79, 91]$ \\
$\mathbf{3}$ & $  24.16$~~$[ 16,  33]$ & $  44.91$~~$[ 33,  56]$ &
$ 138.32$~~$[125, 151]$ & $ 192.61$~~$[177, 207]$ \\
$\mathbf{4}$ & $  20.00$~~$[ 12,  29]$ & $  51.37$~~$[ 40,  64]$ &
$ 191.68$~~$[172, 211]$ & $ 438.95$~~$[418, 460]$ \\
\hline
\end{tabular*}
\end{table}

An even better alternative is to use the whole posterior distribution to
propagate the randomness in $\mathcal{T}$ in our subsequent analyses.
Consider, for instance, the mean regional cost
\[
c(\mathcal{T}) = \sum_{i,j} c_{ij} T_{ij} / T,
\]
and let us compare its posterior distribution, as induced by $\mathcal{T}$, to
the fixed value $C_p$---the mean prior regional cost---we set as a restriction
in \eqref{eq:costconst} to define $\beta$. Since $\beta=0.1$, $C_p = 8.51$.
We can now use our samples $\mathcal{T}^{(1)},\ldots,\mathcal{T}^{(G)}$ from
the Gibbs sampler to generate realizations
\begin{equation}
\label{eq:tripcost}
c(\mathcal{T}^{(g)}) = \sum_{i,j} c_{ij} T_{ij}^{(g)} / T
\end{equation}
and estimate $\Pr(c(\mathcal{T})\,|\,\mathcal{O},\mathcal{D})$.
Figure~\ref{fig:cost1} shows a histogram based on $\{c(\mathcal{T}^{(g)})\}$.
The estimated posterior mean cost is
$\Exp[c(\mathcal{T})\,|\,\mathcal{O},\mathcal{D}] =
c(\overline{\mathcal{T}}) = 8.67$, the posterior mode cost---the Furness
solution cost---is
$c(\hat{\mathcal{T}}) = 8.70$, both higher than $C_p$,
while a $95\%$ credible interval for $c(\mathcal{T})$ is $[8.46, 8.88]$,
barely covering $C_p$; moreover,
\[
\Pr\big( c(\mathcal{T}) \ge C_p\,|\,\mathcal{O},\mathcal{D} \big) \approx
\frac{1}{G}\sum_{g=1}^G I\big[ c(\mathcal{T}^{(g)}) \ge C_p \big] = 0.93.
\]
That a great proportion of possible trip patterns is spending more than
previously expected strongly suggests that a lower value for $\beta$ would be
more realistic given the restrictions on $\mathcal{T}$ by $\mathcal{O}$ and
$\mathcal{D}$.

\begin{figure}[htbp]
\begin{center}
\includegraphics[width=.7\textwidth]{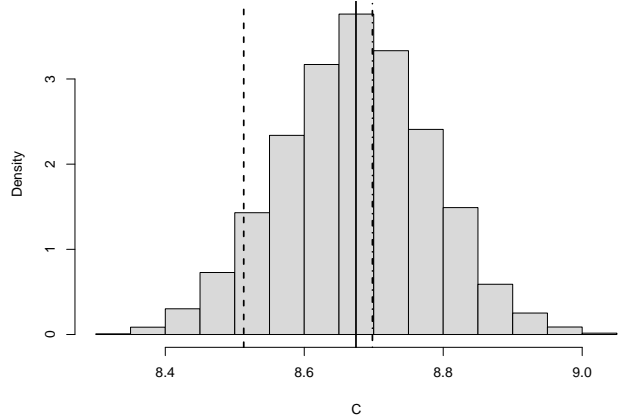}
\end{center}
\caption{Estimated posterior distribution of mean regional cost from
$10,\!000$ samples. Solid line indicates posterior mean, dashed line marks
prior mean, and dash-dotted line marks posterior mode cost.}
\label{fig:cost1}
\end{figure}

We might also want to analyse the trip length distribution (TLD) of the
system: given a set of $K$ cost ranges $(c_0, c_1], \ldots, (c_{K-1}, c_K]$,
where $0 \le c_0 < c_1 < \cdots < c_K < \infty$, we bin the proportion of
trips $T_k / T$ with costs in the $k$-th range $(c_{k-1}, c_k]$ for each
$k=1,\ldots,K$. We again use our samples to generate an estimate for each
$T_k$:
\begin{equation}
\label{eq:tld}
T_k^{(g)} = \sum_{i,j} T_{ij}^{(g)} I\big\{ c_{ij} \in (c_{k-1}, c_k] \big\}.
\end{equation}
Table~\ref{tab:tld1} compares the mean posterior TLD with the prior TLD using
aggregated range proportions $\{p_k\}_{k=1,\ldots,K}$, where
$p_k = \sum_{i,j} p_{ij} I\big\{ c_{ij} \in (c_{k-1}, c_k] \big\}$.
Figure~\ref{fig:tld1} represents both TLD with additional $95\%$ credible
intervals for each range. The discrepancy between prior proportions
$\mathbf{p}$ and posterior proportions $T_{ij}/T$ is now more evident due to
the structure in the TLD. In the next section we will propose a principled way
to narrow the gap between these two regional features.

\begin{table}[bht]
\caption{Mean posterior TLD and prior TLD from proportions $\mathbf{p}$.}
\label{tab:tld1}
\begin{tabular*}{\textwidth}
{c@{\extracolsep{\fill}}c@{\extracolsep{\fill}}c@{\extracolsep{\fill}}c@{\extracolsep{\fill}}c@{\extracolsep{\fill}}c@{\extracolsep{\fill}}c} \hline
Range & $(0, 4]$ & $(4, 8]$ & $(8, 12]$
& $(12, 16]$ & $(16, 20]$ & $(20, 24]$ \\ \hline
$\Exp[T_k/T \,|\,\mathcal{O},\mathcal{D}]$ &
$0.18$ & $0.49$ & $0.08$ & $0.09$ & $0.11$ & $0.05$ \\
$p_k$ &
$0.26$ & $0.38$ & $0.11$ & $0.13$ & $0.08$ & $0.04$ \\
\hline
\end{tabular*}
\end{table}

\begin{figure}[htbp]
\begin{center}
\includegraphics[width=.7\textwidth]{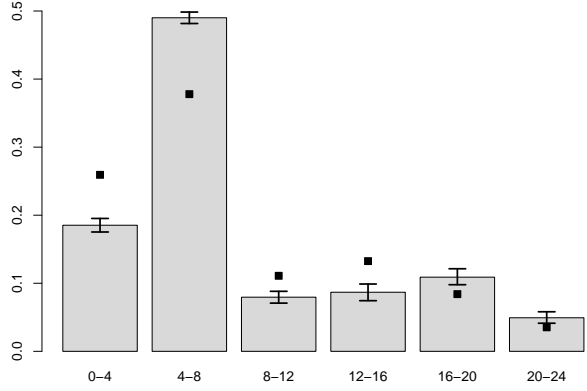}
\end{center}
\caption{Mean posterior TLD (bars) with $95\%$ credible intervals (whiskers),
and prior TLD (squares).}
\label{fig:tld1}
\end{figure}

\end{example}

\section{Extensions to the Proposed Model}
\label{sec:hierarchical}
As we have seen in the last example in the previous section, prior beliefs
might be deceptively outdated or based on regions that are not similar to the
current study region. As a consequence, the related posterior distribution
might be wrongly biased and scaled, affecting the estimation. In addition, it
is possible that during the process of eliciting the prior proportions we
realize that the trip structure in the region is uncertain as it might change
during the study time frame due to, for example, seasonal effects.

A natural approach is then to adopt our same viewpoint with respect to trip
patterns and to explicitly quantify the uncertainty by regarding the
proportions themselves as random, yielding a \emph{hierarchical} model.
Under this updated model the proportions $\mathbf{p}$ are now random and our
samples from the last section are now conditional on $\mathbf{p}$, that is,
$\Pr(\mathcal{T}\,|\,\mathcal{O},\mathcal{D})$
becomes $\Pr(\mathcal{T}\,|\,\mathbf{p},\mathcal{O},\mathcal{D})$.
Nevertheless, we can still proceed in the same way we have done before if we
integrate out the uncertainty in the nuisance parameters, the proportions,
to obtain the marginal posterior distribution on the trips $\mathcal{T}$,
\begin{equation}
\label{eq:hier0}
\Pr(\mathcal{T}\,|\,\mathcal{O}, \mathcal{D})
= \int \Pr(\mathcal{T}, \mathbf{p} \,|\, \mathcal{O}, \mathcal{D})
d\mathbf{p}.
\end{equation}
It is noteworthy that similarly to the previous posterior derivations,
\[
\Pr(\mathcal{T}, \mathbf{p} \,|\, \mathcal{O}, \mathcal{D})
\propto \Pr(\mathcal{O}, \mathcal{D}\,|\,\mathcal{T}, \mathbf{p})
\Pr(\mathcal{T}, \mathbf{p})
= \Pr(\mathcal{O}, \mathcal{D}\,|\,\mathcal{T})
\Pr(\mathcal{T}\,|\, \mathbf{p}) \Pr(\mathbf{p}),
\]
that is, we now simply condition $\mathcal{T}$ on $\mathbf{p}$ (compare with
the numerator in~\eqref{eq:posterior}). The integral in~\eqref{eq:hier0} can
be hard to evaluate directly, but we can again resort to Monte Carlo methods
to sample from
$\Pr(\mathcal{T}\,|\,\mathcal{O},\mathcal{D})$ and conduct the inference, as
we will see shortly.

Even though a hierarchical model increases complexity, it has two main
advantages. First, we can now explain the uncertainty in trip pattern
structure by specifying a suitable probability distribution for $\mathbf{p}$.
This way, lack of information about trip pattern behaviors in the study region
is reflected by more variability in the proportions, which, in turn, results
in more dispersed trip pattern posterior distributions.

Secondly, we can better incorporate additional data that are related to
the trip pattern structure. For instance, if there is available preliminary
data $\mathcal{T}_0$---usually from a small scale study in the same region or
from a region with very similar structure---we can seamlessly incorporate it
in the inference through the posterior
$\Pr(\mathcal{T}\,|\,\mathcal{O},\mathcal{D},\mathcal{T}_0)$. This last
posterior distribution can be obtained by adding the extra conditional on
$\mathcal{T}_0$ in~\eqref{eq:hier0} and defining the likelihood
$\Pr(\mathcal{T}_0\,|\,\mathbf{p})$ to derive
\begin{equation}
\label{eq:hierpost}
\begin{split}
\Pr(\mathcal{T}, \mathbf{p} \,|\, \mathcal{O}, \mathcal{D}, \mathcal{T}_0)
&\propto \Pr(\mathcal{O}, \mathcal{D}, \mathcal{T}_0 \,|\,
\mathcal{T}, \mathbf{p}) \Pr(\mathcal{T}, \mathbf{p})\\
&= \Pr(\mathcal{O}, \mathcal{D}\,|\,\mathcal{T})
\Pr(\mathcal{T}_0 \,|\, \mathbf{p})
\Pr(\mathcal{T} \,|\, \mathbf{p}) \Pr(\mathbf{p}).
\end{split}
\end{equation}
Note that we make the reasonable assumption that $\mathcal{T}$ and
$\mathcal{T}_0$ are conditionally independent given $\mathbf{p}$.

An alternative, common approach is to assume that the proportions $\mathbf{p}$
are unknown, use $\mathcal{T}_0$ to estimate them, and then adopt the obtained
estimate as if it were the ``true'' value of $\mathbf{p}$; this approach is
called \emph{empirical Bayes} in the statistical literature, but is
traditionally referred to as \emph{calibration} in OD matrix estimation.
Albeit being computationally simpler, this treatment has the drawback of
underestimating variance, that is, it does not fully reflect the total
uncertainty in the inference \citep{kasssteffey89}.

To better elucidate the proposed hierarchical models we present two
applications next.

\subsection{Incorporating seed matrices}
A good candidate for the hyperprior distribution on $\mathbf{p}$ is the
multinomial conjugate distribution, the Dirichlet distribution,
$\mathbf{p} \sim \mbox{\texttt{Dir}}(\boldsymbol{\pi})$, with mass function
\[
\Pr(\mathbf{p}) \propto \prod_{i,j} p_{ij}^{\pi_{ij} - 1}.
\]
We then have
\[
\begin{split}
\Pr(\mathcal{T}, \mathbf{p} \,|\, \mathcal{O}, \mathcal{D})
&\propto \prod_{i,j} \frac{p_{ij}^{T_{ij}}}{T_{ij}!}
\prod_{i,j} p_{ij}^{\pi_{ij} - 1}
I[\mathcal{T} \in C(\mathcal{O}, \mathcal{D})]\\
&= \prod_{i,j} \frac{p_{ij}^{T_{ij} + \pi_{ij} - 1}}{T_{ij}!}
I[\mathcal{T} \in C(\mathcal{O}, \mathcal{D})].
\end{split}
\]
A non-informative prior on $\mathbf{p}$ is attained by setting
$\boldsymbol{\pi} = (1, \ldots, 1)$ which is equivalent to $\mathbf{p}$ having
a uniform distribution over all $\{p_{ij}\} \in [0,1]^{n^2}$ such that
$\sum_{i,j} p_{ij} = 1$. In this case, the expression for
$\Pr(\mathcal{T}, \mathbf{p} \,|\, \mathcal{O}, \mathcal{D}, \mathcal{T}_0)$
above is exactly the same as~\eqref{eq:posterior}, but with the important
distinction of now being a joint distribution since $\mathbf{p}$ is random.

Suppose now that we have preliminary data
$\mathcal{T}_0 = \{t_{ij}\}_{i,j=1,\ldots,n}$ in the form of a seed matrix of
trip counts. In the classical approach discussed in the introduction,
$\mathcal{T}_0$ is commonly used to estimate the proportions as
$\hat{p}_{ij} = t_{ij} / T_0$, where $T_0 = \sum_{k,l} t_{kl}$, or to simply
kick-start an estimation procedure. This approach, however, effectively
ignores the sample size $T_0$ since $\hat{p}_{ij}$ remains the same if we
observe $\kappa$ times more counts, $\kappa \mathcal{T}_0$, even for $\kappa$
arbitrarily large; furthermore, similarly to empirical Bayes, it yields lower
posterior variances for $\mathcal{T}$.

Following our discussion, here we offer a more principled way to
incorporate the seed matrix $\mathcal{T}_0$ by performing posterior inference
on $\mathcal{T}$ through the distribution in~\eqref{eq:hierpost}. We assume
that, similar to $\mathcal{T}$, the seed counts follow a conditional
multinomial distribution, $\mathcal{T}_0 \sim \texttt{MN}(T_0, \mathbf{p})$
with flat prior $\Pr(T_0) \propto 1$. Adopting the same Dirichlet distribution
for $\mathbf{p}$ we have
\begin{equation}
\begin{split}
\Pr(\mathcal{T}, \mathbf{p} \,|\, \mathcal{O}, \mathcal{D}, \mathcal{T}_0)
&\propto \prod_{i,j} \frac{p_{ij}^{T_{ij}}}{T_{ij}!}
\prod_{i,j} \frac{p_{ij}^{t_{ij}}}{t_{ij}!}
\prod_{i,j} p_{ij}^{\pi_{ij} - 1}
I[\mathcal{T} \in C(\mathcal{O}, \mathcal{D})] \\
&\propto \prod_{i,j} \frac{p_{ij}^{T_{ij} + t_{ij} + \pi_{ij} - 1}}{T_{ij}!}
I[\mathcal{T} \in C(\mathcal{O}, \mathcal{D})],
\end{split}
\end{equation}
and thus
$\mathbf{p}\,|\,\mathcal{T}, \mathcal{T}_0 \sim
\mbox{\texttt{Dir}}(\boldsymbol{\pi} + \mathcal{T} + \mathcal{T}_0)$.

To sample from
$\Pr(\mathcal{T}, \mathbf{p} \,|\, \mathcal{O}, \mathcal{D}, \mathcal{T}_0)$
we adopt an extended Gibbs sampler with an extra step that accommodates the
new hierarchical level: we iteratively sample from
$\Pr(\mathcal{T}\,|\, \mathbf{p}, \mathcal{O}, \mathcal{D}, \mathcal{T}_0)
= \Pr(\mathcal{T}\,|\, \mathbf{p}, \mathcal{O}, \mathcal{D})$ exactly how we
were doing in the previous section, and sample from the conditional Dirichlet
$\Pr(\mathbf{p}\,|\, \mathcal{T}, \mathcal{O}, \mathcal{D}, \mathcal{T}_0)
= \Pr(\mathbf{p}\,|\, \mathcal{T}, \mathcal{T}_0)$. If a seed matrix is not
available, the second step becomes simply sampling from $\Pr(\mathbf{p}\,|\,
\mathcal{T})$, still a Dirichlet distribution. The updated Gibbs sampler is
listed below.

\begin{enumerate}[Step 1.]
\item Start at some arbitrary initial configuration $\mathcal{T}^{(0)}$ and
initial proportions $\mathbf{p}^{(0)}$.
\item For $t = 1, 2, \ldots$ do (until convergence):
  \begin{enumerate}[Step 2.1.]
  \item For $i,j = 1, \ldots, n-1$ do:
  sample
  $T_{ij}^{(t)} \sim T_{ij}\,|\,T_{[ij]}^{(t-1)}, \mathbf{p}^{(t-1)},
  \mathcal{O}, \mathcal{D}$ from a non-central hypergeometric using a
  Metropolis step,
  \[
  T_{ij}^{(t)} =
  MS(T_{ij}^{(t-1)}; O_{ij}^{(t-1)}, D_{ij}^{(t-1)}, \Delta_{ij}^{(t-1)},
  \psi_{ij}^{(t-1)}),
  \]
  with $O_{ij}^{(t-1)}$, $D_{ij}^{(t-1)}$, and $\Delta_{ij}^{(t-1)}$ as
  before, and
  \[
  \psi_{ij}^{(t-1)} = (p_{ij}^{(t-1)} p_{nn}^{(t-1)})
  / (p_{in}^{(t-1)} p_{nj}^{(t-1)}). 
  \]

  \item Sample
  $\mathbf{p}^{(t)} \sim \mbox{\texttt{Dir}}(\mathcal{T}^{(t)} + \mathcal{T}_0
  + \boldsymbol{\pi})$
  or $\mathbf{p}^{(t)} \sim \mbox{\texttt{Dir}}(\mathcal{T}^{(t)}
  + \boldsymbol{\pi})$ if $\mathcal{T}_0$ is not available.
  \end{enumerate}
\end{enumerate}

To perform inference on the marginal posterior
$\Pr(\mathcal{T}\,|\,\mathcal{O}, \mathcal{D}, \mathcal{T}_0)$ we just need to
use the realizations from the Gibbs sampler; the posterior mean, for instance,
is readily available from~\eqref{eq:posteriormean}. MAP estimates, however,
are harder to obtain since we need to compute the integral
in~\eqref{eq:hier0}. One alternative is to use the joint posterior mode,
\[
\tilde{\mathcal{T}} = \argmax_{\mathcal{T} \in C(\mathcal{O}, \mathcal{D})}
\Bigg\{ \max_{\mathbf{p} \in [0,1]^{n^2} : \sum_{i,j} p_{ij} = 1}
\Pr(\mathcal{T}, \mathbf{p}\,|\,\mathcal{O}, \mathcal{D}, \mathcal{T}_0)
\Bigg\},
\]
but then the estimate might be biased since it is conditional on the optimal
value of $\mathbf{p}$. In the same vein, we could first ``calibrate'' by
setting some specific $\mathbf{p}$, say the marginal posterior mean
\[
\overline{\mathbf{p}} = \Exp[\mathbf{p}\,|\,\mathcal{O}, \mathcal{D},
\mathcal{T}_0] \approx \frac{1}{G}\sum_{g=1}^G \mathbf{p}^{(g)},
\]
and then produce
\begin{equation}
\label{eq:condmap}
\hat{\mathcal{T}} = \argmax_{\mathcal{T} \in C(\mathcal{O}, \mathcal{D})}
\Pr(\mathcal{T}\,|\,\overline{\mathbf{p}}, \mathcal{O}, \mathcal{D},
\mathcal{T}_0).
\end{equation}
It can be shown that the first estimator, $\tilde{\mathcal{T}}$, can be
obtained by an extended Furness method that iteratively solves for
$\mathbf{p}$ while fitting the balancing factors by setting
\[
\tilde{p}_{ij} = \frac{\tilde{T}_{ij} + t_{ij} + \pi_{ij} - 1}
{\sum_{k,l=1,\ldots,n} \tilde{T}_{kl} + t_{kl} + \pi_{kl} - 1},
\]
but we will not pursue it further here.

\subsection{Incorporating prior trip length distributions}
Seed matrices provide information on each OD pair in the system and thus
derive more accurate trip pattern inferences. More often than not, however, we
do not have preliminary data $\mathcal{T}_0$ at this level of detail at our
disposal. In some cases $\mathcal{T}_0$ contains censored observations; we
might observe trips in a survey, but these trips are known only to have come
from a certain origin, or to a destination, or to have had some specific
travel cost. For instance, recalling the trip length distribution (TLD) from
Example~2, we might only discriminate a trip in our survey by specifying its
cost ``bin'', that is, within which range its cost falls.

Assume that we know the OD trip costs $\{c_{ij}\}$ and consider, as
before, the $K$ cost ranges $(c_0, c_1], \ldots, (c_{K-1}, c_K]$. Our
preliminary counts now fall into $K$ possible strata, $\mathcal{T}_0 = \{t_1,
\ldots, t_K\}$, depending on their transport costs: we observe $t_1$ trips
with costs between $c_0$ and $c_1$, $t_2$ trips spending between and $c_1$ and
$c_2$, and so on. If we again define range proportions aggregated by cost
$\mathbf{p}_0 = \{p_k\}_{k=1\,\ldots,K}$, where
$p_k = \sum_{i,j} p_{ij} I\{c_{ij} \in (c_{k-1}, c_k]\}$, we can then
analogously set
$\mathcal{T}_0\,|\,\mathbf{p} \sim \texttt{MN}(T_0, \mathbf{p}_0)$
with $\Pr(T_0) \propto 1$ as the preliminary data likelihood. We note that
$\mathbf{p}_0$ is a function of $\mathbf{p}$.

We can assume the same Dirichlet distribution for the proportions,
$\mathbf{p} \sim \mbox{\texttt{Dir}}(\boldsymbol{\pi})$, but since
\[
\Pr(\mathcal{T}, \mathbf{p} \,|\, \mathcal{O}, \mathcal{D}, \mathcal{T}_0)
\propto \prod_{i,j} \frac{p_{ij}^{T_{ij}}}{T_{ij}!}
\prod_{k} \frac{p_k^{t_k}}{t_k!}
\prod_{i,j} p_{ij}^{\pi_{ij} - 1}
I[\mathcal{T} \in C(\mathcal{O}, \mathcal{D})]
\]
and each $p_k$ is a sum of $p_{ij}$ for all pairs $i$ and $j$ with cost in the
$k$-th bin, we lose the conjugacy. Another approach, in case we are more
informed about the censored proportions, is to opt for a Dirichlet prior on
$\mathbf{p}_0$; but then we again lack conjugacy. Regardless, we can still
obtain a Gibbs sampler that is very similar to the scheme shown in the
previous subsection; we just need to substitute the direct Dirichlet sampling
step, Step~2.2, by another Metropolis step. Next, we provide an updated
sampling scheme in a simpler context.

Suppose that the proportions follow a gravity model with $p_{ij} \propto
\exp(-\beta c_{ij})$, as in the previous section, but now we make $\beta$
random to drive the uncertainty in $\mathbf{p}$. Moreover, we settle on a
Dirichlet prior on $\mathbf{p}_0$,
$\mathbf{p}_0(\beta) \sim \mbox{\texttt{Dir}}(\boldsymbol{\pi})$,
where $\boldsymbol{\pi} = \{\pi_1, \ldots, \pi_K\}$.
In what follows we explicitly represent the dependency of the proportions on
$\beta$ for clarity; we also note that now
\[
p_k(\beta) \propto \sum_{i,j} \exp(-\beta c_{ij})
I\{c_{ij} \in (c_{k-1}, c_k]\}.
\]

The joint posterior is thus given by
\begin{equation}
\label{eq:jointbeta}
\begin{split}
\Pr(\mathcal{T}, \beta \,|\, \mathcal{O}, \mathcal{D}, \mathcal{T}_0)
&\propto \prod_{i,j} \frac{p_{ij}(\beta)^{T_{ij}}}{T_{ij}!}
\prod_{k} \frac{p_k(\beta)^{t_k}}{t_k!}
\prod_{k} p_k(\beta)^{\pi_k - 1}
I[\mathcal{T} \in C(\mathcal{O}, \mathcal{D})] \\
&\propto \underbrace{\prod_{i,j} p_{ij}(\beta)^{T_{ij}}
\prod_k p_k(\beta)^{t_k + \pi_k - 1}}_{\Phi(\beta; \mathcal{T}, \mathcal{T}_0)}
I[\mathcal{T} \in C(\mathcal{O}, \mathcal{D})]. \\
\end{split}
\end{equation}
From~\eqref{eq:jointbeta} we deduce that setting
$\boldsymbol{\pi} = \{1, \ldots, 1\}$ for a non-informative Dirichlet prior is
equivalent to having a flat improper prior for the cost deterrence,
$\Pr(\beta) \propto 1$.

The Gibbs sampler has two iterative steps: we alternate between sampling from
$\mathcal{T}$ conditional on the impedance $\beta$ and all the data,
$\Pr(\mathcal{T}\,|\,\beta,\mathcal{O},\mathcal{D},\mathcal{T}_0)$, and
sampling from $\beta$ conditional on trip patterns $\mathcal{T}$ and margins
and preliminary data,
$\Pr(\beta\,|\,\mathcal{T},\mathcal{O},\mathcal{D},\mathcal{T}_0)$. We already
know, since Section~\ref{sec:model}, how to sample from
$\Pr(\mathcal{T}\,|\,\beta,\mathcal{O},\mathcal{D},\mathcal{T}_0)
= \Pr(\mathcal{T}\,|\,\mathbf{p}(\beta),\mathcal{O},\mathcal{D})$
using random walk Metropolis steps for the non-central hypergeometric.
To sample from
$\Pr(\beta\,|\,\mathcal{T},\mathcal{O},\mathcal{D},\mathcal{T}_0)$ we
construct another random walk Metropolis step.

First, let us define the normalizing factors
$Z_k(\beta) = \sum_{i,j} \exp(-\beta c_{ij}) I\{c_{ij} \in (c_{k-1}, c_k]\}$
and $Z(\beta) = \sum_{i,j} \exp(-\beta c_{ij}) = \sum_k Z_k(\beta)$, so that
$p_{ij} = \exp(-\beta c_{ij})/Z(\beta)$ and $p_k = Z_k(\beta)/Z(\beta)$. Also,
recall that $T = \sum_{i,j} T_{ij}$, $T_0 = \sum_k t_k$, and define
$T_0^* = \sum_k (t_k + \pi_k - 1) = T_0 + \sum_k \pi_k - K$.
The function $\Phi(\beta; \mathcal{T}, \mathcal{T}_0)$ in the joint
posterior~\eqref{eq:jointbeta} then simplifies to
\[
\begin{split}
\Phi(\beta; \mathcal{T}, \mathcal{T}_0) &=
\prod_{i,j} \Bigg(\frac{\exp(-\beta c_{ij})}{Z(\beta)}\Bigg)^{T_{ij}}
\prod_k \Bigg(\frac{Z_k(\beta)}{Z(\beta)}\Bigg)^{t_k + \pi_k - 1} \\
&= \exp\Bigg\{
-\beta \sum_{i,j} c_{ij} T_{ij} + \sum_k (t_k + \pi_k - 1) \log Z_k(\beta)\\
&\qquad\mbox{}-(T + T_0^*) \log Z(\beta) \Bigg\}.
\end{split}
\]
As proposal distribution, let us select a normal distribution centered at the
current realization of $\beta$ in the chain with small variance $\sigma^2$. To
get $\beta^{(t)}$ at the $t$-th iteration we then sample a candidate
$\beta^* \sim N(\beta^{(t-1)}, \sigma^2)$ and accept or reject it based on the
acceptance ratio
\begin{equation}
\label{eq:Rbeta}
R(\beta^{(t-1)}, \beta^*) =
\frac{\Pr(\beta^*\,|\,\mathcal{T},\mathcal{O},\mathcal{D},\mathcal{T}_0)}
{\Pr(\beta^{(t)}\,|\,\mathcal{T},\mathcal{O},\mathcal{D},\mathcal{T}_0)}
= \frac{\Phi(\beta^*; \mathcal{T}^{(t-1)}, \mathcal{T}_0^{(t-1)})}
{\Phi(\beta^{(t-1)}; \mathcal{T}^{(t-1)}, \mathcal{T}_0^{(t-1)})}.
\end{equation}
The final, updated Gibbs sampler is listed below.

\begin{enumerate}[Step 1.]
\item Start at some arbitrary initial configuration $\mathcal{T}^{(0)}$ and
initial impedance $\beta^{(0)}$.
\item For $t = 1, 2, \ldots$ do (until convergence):
  \begin{enumerate}[Step 2.1.]
  \item For $i,j = 1, \ldots, n-1$ do:
  sample
  $T_{ij}^{(t)} \sim T_{ij}\,|\,T_{[ij]}^{(t-1)}, \mathbf{p}(\beta^{(t-1)}),
  \mathcal{O}, \mathcal{D}$ from a non-central hypergeometric using a
  Metropolis step,
  \[
  T_{ij}^{(t)} =
  MS(T_{ij}^{(t-1)}; O_{ij}^{(t-1)}, D_{ij}^{(t-1)}, \Delta_{ij}^{(t-1)},
  \psi_{ij}(\beta^{(t-1)})).
  \]
  with
  \[
  \psi_{ij}(\beta^{(t-1)}) =
  \frac{p_{ij}(\beta^{(t-1)}) p_{nn}(\beta^{(t-1)})}
  {p_{in}(\beta^{(t-1)}) p_{nj}(\beta^{(t-1)})}.
  \]

  \item Sample candidate $\beta^* \sim N(\beta^{(t-1)}, \sigma^2)$ and set
  $\beta^{(t)} = \beta^*$ (accept) with probability $\min\{1, R(\beta^{(t-1)},
  \beta^*)\}$ where $R(\cdot)$ is the ratio in~\eqref{eq:Rbeta}; otherwise,
  set $\beta^{(t)} = \beta^{(t-1)}$ (reject.)

  \end{enumerate}
\end{enumerate}

\begin{exrev}
Under the same setting of Example~2, but now with $\beta$ random, let us
initially set $\boldsymbol{\pi} = \{1,\ldots,1\}$, that is, a non-informative
prior on $\beta$. We run a Gibbs sampler with proposal variance $\sigma^2 =
10^{-4}$ until convergence and take $G = 10,\!000$ samples for posterior
inference.

Our estimate for $\beta$,
$
\overline{\beta} = \Exp[\beta\,|\,\mathcal{O}, \mathcal{D}]
\approx \frac{1}{G} \sum_{g=1}^G \beta^{(g)} = 0.031,
$
is much lower than the assumed value in Example~2 ($\beta = 0.1$), which
corroborates with our previous remark about a more realistic value for the
cost impedance. Such lower values are expected since the inference is solely
driven by the observed data and thus better represents the margin constraints.
The estimated $95\%$ credible interval for $\beta$ is large, $[0.009, 0.056]$,
reflecting the high degree of uncertainty that arises from trying to capture
the structural trip proportions using a single parameter.

The effect of a random $\beta$ in trip patterns can be appreciated in the
estimated marginal posterior distributions for $\mathcal{T}$ pictured in
Figure~\ref{fig:exrev0T}. We draw attention to the increased spread when
compared to the distributions in Figure~\ref{fig:boxplot1}. We also observe
that the Furness solution, conditional on $\overline{\beta}$ and represented
by squares, is similar to the posterior mean
$\Exp[\mathcal{T}\,|\,\mathcal{O},\mathcal{D}]$.

\begin{figure}[htbp]
\begin{center}
\includegraphics[width=.7\textwidth]{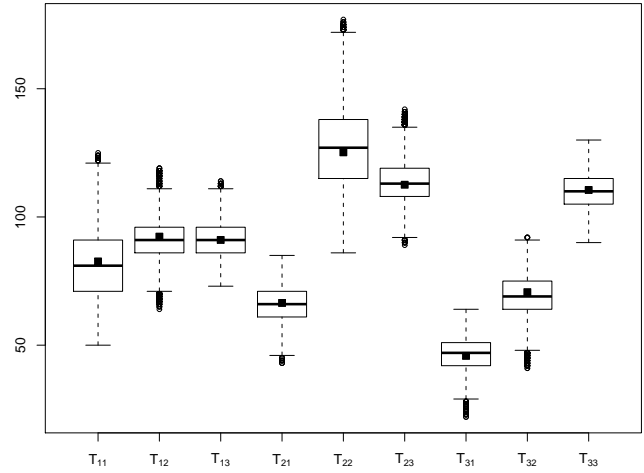}
\end{center}
\caption{Estimated marginal posterior distributions for $\mathcal{T}$ from
hierarchical model with non-informative prior on $\beta$. Squares mark
conditional Furness solution.}
\label{fig:exrev0T}
\end{figure}

%> sum(tm*C)/T
%[1] 11.01345
%> lci(alpha)(spp)
%[1] 10.08614
%> uci(alpha)(spp)
%[1] 11.82161
%> mean(sp)
%[1] 11.01042
%> sum(f*C)/T
%[1] 11.00692

The higher variability in $\mathcal{T}$ is reproduced by wider credible
intervals in the trip length distribution, as shown in
Figure~\ref{fig:exrev0tld}: each bar represents the estimated posterior mean
of $T_k/T$ for each cost range, the squares pinpoint the posterior mean of
$p_k(\beta)$, while the dotted line corresponds to the prior mean $1/K$. As
can be seen, the dependence of the proportions on a single parameter makes the
distribution on $\mathbf{p}$ not flexible enough to follow $\mathcal{T}$
closely. We note again the higher variability in the posterior TLD as assessed
by the wider $95\%$ credible intervals (whiskers) when compared to
Figure~\ref{fig:tld1}.

\begin{figure}[htbp]
\begin{center}
\includegraphics[width=.7\textwidth]{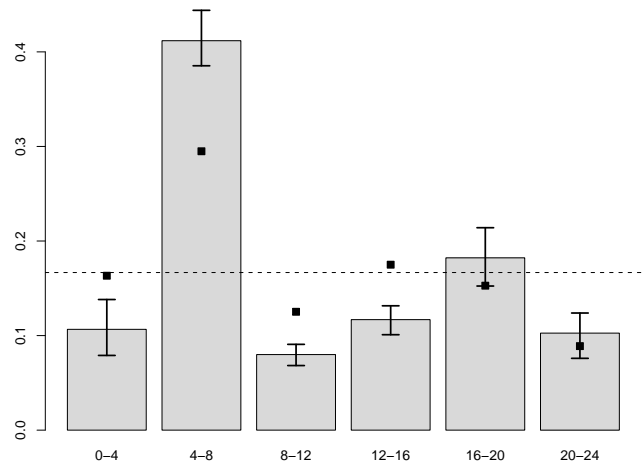}
\end{center}
\caption{Mean posterior TLD (bars) with $95\%$ credible intervals (whiskers),
and mean posterior TLD proportions (squares). The dotted line marks the prior
mean, $1/K$.}
\label{fig:exrev0tld}
\end{figure}

Suppose now that we observe preliminary data $\mathcal{T}_0$ from
\citep[pg.~186]{ortuzarwillusen} in Table~\ref{tab:exrevT0}. Keeping the flat
prior on $\beta$ and $\sigma^2 = 10^{-4}$, we perform posterior inference from
$10,\!000$ samples taken from the Gibbs sampler after convergence.

\begin{table}[bht]
\caption{Preliminary TLD. Data reproduced from
\citep[table~5.14]{ortuzarwillusen}.}
\label{tab:exrevT0}
\begin{tabular*}{\textwidth}
{c@{\extracolsep{\fill}}c@{\extracolsep{\fill}}c@{\extracolsep{\fill}}c@{\extracolsep{\fill}}c@{\extracolsep{\fill}}c@{\extracolsep{\fill}}c} \hline
Range & $(0, 4]$ & $(4, 8]$ & $(8, 12]$
& $(12, 16]$ & $(16, 20]$ & $(20, 24]$ \\ \hline
$t_k$ &
$365$ & $962$ & $160$ & $150$ & $230$ & $95$ \\
$t_k/T_0$ &
$0.19$ & $0.49$ & $0.08$ & $0.08$ & $0.12$ & $0.05$ \\
\hline
\end{tabular*}
\end{table}

The preliminary TLD counts are very informative, $T_0 = T = 1962$, and greatly
affect the inference: our updated estimate for the cost deterrence is a higher
$\overline{\beta} = \Exp[\beta\,|\,\mathcal{O},\mathcal{D},\mathcal{T}_0] =
0.086$, closer to the original $\beta = 0.1$ in Example~2, and the $95\%$
credible interval for $\beta$ is much tighter, $[0.086, 0.093]$. 

The posterior inference on trip patterns is summarized by
Table~\ref{tab:exrevT}, showing posterior mean $\overline{\mathcal{T}}$ and
marginal $95\%$ credible intervals, and Figure~\ref{fig:exrevT}. The marginal
distributions have increased variability when compared to Example~2 due to the
randomness in the proportions, as expected. The variance is, however, not much
higher since the preliminary TLD is very informative. The conditional Furness
solution $\hat{\mathcal{T}}$, shown in square marks in
Figure~\ref{fig:exrevT}, is very similar to the posterior mean. The estimated
posterior probabilities of these solutions are
$\Pr(\overline{\mathcal{T}}\,|\,
\overline{\beta},\mathcal{O},\mathcal{D},\mathcal{T}_0) = 1.3 \cdot 10^{-3}$
and
$\Pr(\hat{\mathcal{T}}\,|\,
\overline{\beta},\mathcal{O},\mathcal{D},\mathcal{T}_0) = 1.5 \cdot 10^{-3}$,
slightly smaller than in Example~2.

\begin{table}[bht]
\caption{Marginal posterior mean and $95\%$ credible intervals.}
\label{tab:exrevT}
\begin{tabular*}{\textwidth}
{c@{\extracolsep{\fill}}c@{\extracolsep{\fill}}c@{\extracolsep{\fill}}c@{\extracolsep{\fill}}c} \hline
Zone & $\mathbf{1}$ & $\mathbf{2}$ & $\mathbf{3}$ & $\mathbf{4}$ \\ \hline
$\mathbf{1}$ & $141.34$~~$[128, 155]$ & $101.49$~~$[87, 118]$ & $71.11$~~$[57, 85]$ & $86.07$~~$[71, 103]$ \\
$\mathbf{2}$ & $63.87$~~$[52, 76]$ & $184.96$~~$[168, 204]$ & $106.10$~~$[89, 120]$ & $105.07$~~$[90, 122]$ \\
$\mathbf{3}$ & $28.47$~~$[20, 37]$ & $51.32$~~$[39, 63]$ & $131.06$~~$[116, 146]$ & $189.14$~~$[172, 205]$ \\
$\mathbf{4}$ & $26.31$~~$[17, 37]$ & $62.23$~~$[48, 77]$ & $191.73$~~$[174, 209]$ & $421.72$~~$[400, 444]$ \\
\hline
\end{tabular*}
\end{table}

\begin{figure}[htbp]
\begin{center}
\includegraphics[width=.7\textwidth]{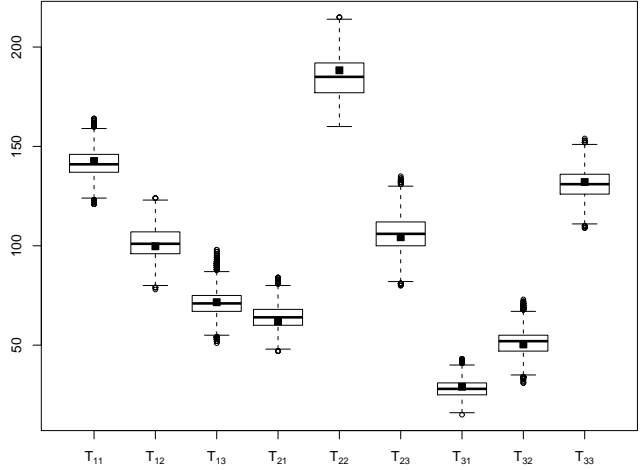}
\end{center}
\caption{Estimated marginal posterior distributions for $\mathcal{T}$ from
hierarchical model. Squares mark conditional Furness solution.}
\label{fig:exrevT}
\end{figure}

Since $\beta < 0.1$ with high posterior probability, we should expect the
system to spend more when compared to the scenario in Example~2.
Figure~\ref{fig:exrevcost} displays the posterior distribution of trip
costs $c(\mathcal{T})$, as estimated from~\eqref{eq:tripcost}. The posterior
mean regional cost
$c(\overline{\mathcal{T}}) =
\Exp[c(\mathcal{T})\,|\,\mathcal{O},\mathcal{D},\mathcal{T}_0]$ is $9.12$,
with a $95\%$ credible interval of $[8.81, 9.45]$, higher than before. The
posterior mode cost $c(\hat{\mathcal{T}})$ is $9.09$, close to
$c(\overline{\mathcal{T}})$, as expected since the estimates are similar. The
proportion cost $C_p(\beta) = \sum_{i,j} c_{ij} p_{ij}(\beta)$
in~\eqref{eq:costconst} inherits the randomness from $\beta$; its estimated
posterior mean, $8.95$, is lower than $c(\overline{\mathcal{T}})$, which can
also be attributed to the rigidness in $\mathbf{p}$.

\begin{figure}[htbp]
\begin{center}
\includegraphics[width=.7\textwidth]{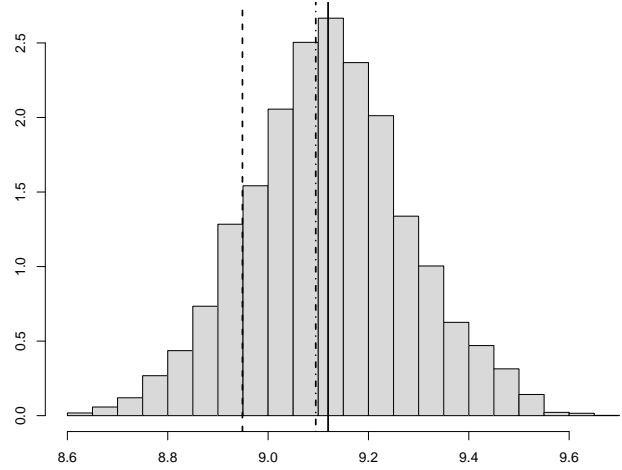}
\end{center}
\caption{Estimated posterior distribution of mean regional cost. Solid line
indicates posterior mean, dashed line marks posterior mean proportion cost,
and dash-dotted line marks posterior mode cost.}
\label{fig:exrevcost}
\end{figure}

Finally, we can also see the effect of $\mathcal{T}_0$ in reducing the
inferential uncertainty in the posterior TLD at Figure~\ref{fig:exrevtld}, as
illustrated by the tighter $95\%$ credible intervals. We still see the
discrepancy between the posterior TLD---whose mean
$\Exp[T_k/T\,|\,\mathcal{O},\mathcal{D},\mathcal{T}_0]$ is
represented by bars---and the posterior proportion TLD---whose mean
$\Exp[p_k(\beta)\,|\,\mathcal{O},\mathcal{D},\mathcal{T}_0]$
is identified by squares. We note, however, that the posterior mean TLD is
close to the prior mean TLD, $t_k/T_0$, represented by diamonds and listed
in Table~\ref{tab:exrevT0}, since $\mathcal{T}_0$ is highly informative and
thus influential. The two mean posterior TLD are listed in
Table~\ref{tab:exrevtld}. 

\begin{figure}[htbp]
\begin{center}
\includegraphics[width=.7\textwidth]{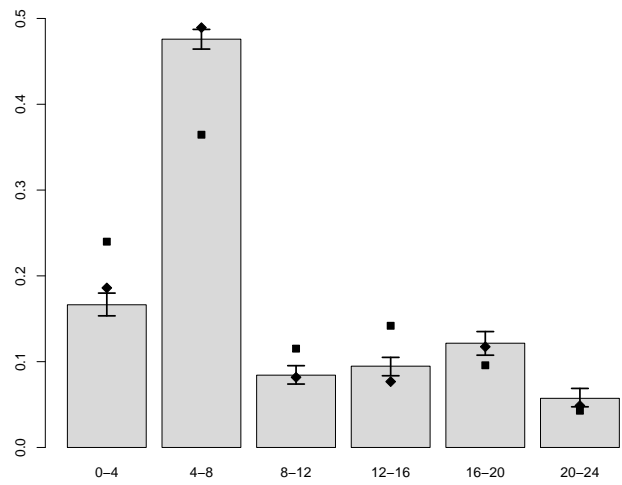}
\end{center}
\caption{Posterior mean TLD (bars) with $95\%$ credible intervals (whiskers),
posterior mean proportion TLD (squares), and prior mean TLD (diamonds).}
\label{fig:exrevtld}
\end{figure}

\begin{table}[bht]
\caption{Posterior mean trip length distributions based on $\mathcal{T}$ and
$\mathbf{p}$.}
\label{tab:exrevtld}
\begin{tabular*}{\textwidth}
{c@{\extracolsep{\fill}}c@{\extracolsep{\fill}}c@{\extracolsep{\fill}}c@{\extracolsep{\fill}}c@{\extracolsep{\fill}}c@{\extracolsep{\fill}}c} \hline
Range & $(0, 4]$ & $(4, 8]$ & $(8, 12]$
& $(12, 16]$ & $(16, 20]$ & $(20, 24]$ \\ \hline
$\Exp[T_k/T\,|,\mathcal{O},\mathcal{D},\mathcal{T}_0]$ &
$0.17$ & $0.48$ & $0.08$ & $0.09$ & $0.12$ & $0.06$ \\
$\Exp[p_k(\beta)\,|,\mathcal{O},\mathcal{D},\mathcal{T}_0]$ &
$0.24$ & $0.36$ & $0.12$ & $0.14$ & $0.10$ & $0.04$ \\
\hline
\end{tabular*}
\end{table}

\end{exrev}

\section{Discussion}
\label{sec:discussion}
Static origin-destination matrix estimation has been traditionally regarded as
an optimization problem. Here we draw from the contingency table literature
and cast OD matrix estimation as a formal statistical inference problem and
adopt a Bayesian approach where trip patterns are considered random.
Furthermore, we make model assumptions on the parameters describing the
probability distribution on trip patterns---trip proportions that govern the
structure of trip distribution---as opposed to the classical assumptions on
particular objective functions. The use of trip proportions frees us from
requiring seemingly artificial constraints on trip configurations, provides
more easily interpretable results, and allows us to better incorporate other
sources of data in a principled way within a Bayesian framework.

By electing specific functional forms for the trip proportions---as based on
the entropy maximizing principle, for example---we are able to recover
classical solutions as MAP estimators and thus inherit the justifications and
rich history behind traditional approaches. Yet, perhaps the main benefit of
our proposed approach is to better characterize the uncertainty in the
solutions and, in general, in trip distribution. As we have showed in many
examples, it is common for any point estimate---such as the Furness solution
or posterior mean---to capture only a small fraction of possible trip
configurations given the large number of alternatives. Point estimators, when
seen as ensemble summarizers, can be useful for preliminary planning purposes
and gaining insight on the trip distribution in the study region; they can,
however, be poor substitutes of the full posterior distribution in further
analyses as they can dramatically underestimate the variability in trip
patterns.

Preliminary data is traditionally used to calibrate specific parameters of the
trip distribution model, such as cost deterrence. Nonetheless, fixing an
optimal data fitting value for the parameter can further underestimate
variance in the inference. In our fully Bayesian approach we explicitly
acknowledge the uncertainty in the parameters by also making them random: we
set a hyper-prior distribution on trip proportions to build a hierarchical
model. As a consequence, and in contrast with a traditional approach, more
informative preliminary data---for example, high counts in a seed
matrix---yield more precise inference on trip configurations as we are able to
more accurately characterize trip proportions.

The adoption of a Bayesian framework carries many other benefits not covered
here: besides point and interval inference, we are also able to test
hypotheses by explicitly comparing models through Bayes factors; moreover,
Bayesian methods can be further explored to perform model validation through
posterior predictive checks. In summary, the flexibility of Bayesian
statistics is particularly helpful and really comes to bear when exploring
high-dimensional spaces such as the ensemble of feasible trip configurations.

There is, however, a price to pay for such modeling power in higher
computational costs, and thus the procedures discussed here still need to be
more closely examined in this respect. Specifically, the increased
complexity in generating and analysing trip configuration samples instead of
simply obtaining the most likely trip assignment needs to be assessed as the
proposed routines are tried in real-world datasets comprising large systems.
Future directions would also include the development of more efficient
sampling schemes through improved algorithms---better proposal densities, for
example---and faster implementations that would explore, for instance,
parallel versions of the proposed procedures.

Finally, it should be noted that the models proposed here can serve as basis
for an integrated higher level model that incorporates other traffic modeling
steps; as an example, the effect of congested networks could be considered in
OD matrix estimation if our model would jointly consider trip distribution and
route assignment. As it is common in Bayesian modeling, we would then be able
to propagate the uncertainty across steps while performing marginal inference
on any aspect of the higher model conditional on data from all steps.
Furthermore, other types of data could also be considered to obtain more
refined models with, for instance, link count data and camera sensors or
temporal variation for dynamic OD matrix estimation.

\section*{Acknowledgements}
The author would like to thank Prof. Felipe Loureiro from Federal University
of Cear\'{a}, Brazil, for many fruitful discussions and a constant source of
motivation.

\bibliographystyle{model2-names}
\bibliography{od}

\end{document}